JTLU

# The Cost of Connectivity: Network Tradeoffs in Subway Expansion and Historic Street Preservation

**Liang, Xiaofan,** University of Michigan – Ann Arbor, xfliang@umich.edu
**Chen, Lu,** Georgia Institute of Technology, chen.lu.g50@gmail.com
**Lyu, Manying,** Guangzhou University, 1072712342@qq.com
**Tian, Yun,** Guangzhou University, 549447124@qq.com
**Ye, Changdong**, South China Agricultural University, yechangdong@scau.edu.cn

**Abstract:** As cities become increasingly interconnected, expanding regional mobility networks can come at the cost of local social networks. This study examines such tradeoffs in the case of Miaoqianzhijie, a historic street in Guangzhou, China, which was planned to be demolished for building a new interchange subway at Line 10. We introduce a network tradeoff framework that analyzes subway lines as spatial networks and historic streets as social networks within a single analytic frame. Empirically, we combine a participatory GIS–based social network survey with origin–destination (OD) flow analysis of Guangzhou's subway system. Our findings show that Miaoqianzhijie functions as a dense corridor of social infrastructure that support everyday routines, social ties, mutual aid, and symbolic place attachment for residents, workers, and frequent visitors. Our OD analysis found that Line 10's mobility benefits are unevenly distributed and the station closest to the demolition area sees only modest improvements. These findings reveal an asymmetric network tradeoff that highlights concentrated social connectivity losses versus diffuse and limited local mobility gains. This study contributes a case study and network-based methodology for representing such tradeoffs and highlights how these representations can inform more equitable planning debates around subway expansion and historic street preservation.

## 1 Introduction

Subway systems are critical transportation infrastructure to support mobility flows in metropolises. The expansion of subway systems is often considered a "public good", as it increases a region's spatial connectivity, accessibility, and polycentricity (Song et al., 2023), stimulates job growth (Jin & Kim, 2018), alleviates spatial mismatch between job and housing, and raises property values for property owners nearby (Bajic, 1983; Agostini & Palmuci, 2008; Jiang & Levinson, 2017; Costa et al., 2022). Within this dominant framing, the primary question is how to extend and optimize the network so that more people can move more efficiently.

Network analysis has supported subway planning by modeling how new lines or stations alter origin-destination flows (Kim & Song, 2015), identifying system bottlenecks (Feng et al., 2017), and diverting traffic (Derrible, 2012). These approaches, often grounded in systems engineering, focus on optimizing spatial connectivity within the subway network. However, they emphasize regional priorities, such as flow efficiency, that may differ from the needs of residents at construction sites. However, these approaches take the regional network as the main unit of analysis and thus foreground system priorities, such as efficiency, resilience, coverage, that may diverge from the interests and experiences of residents at specific construction sites.

These tensions become particularly pronounced when expanding subway infrastructure in densely populated areas requires the demolition of surface buildings. Existing place-based assessments focus on "social costs" such as physical displacement, environmental and quality-of-life impacts, and, to a lesser extent, gentrification (Meyer, 2016, pp.139-146; Lin et al., 2022). Yet, the loss of lived experience, memory, and social ties, are frequently overlooked or reduced to counts of affected households or designated heritage buildings. Viewed through a network lens, these are not only land-use or displacement issues but also conflicts between different kinds of connectivity: regional mobility flows supported by subway networks and local social networks anchored in historic streets.

This paper introduces a network tradeoff framework to highlight the implicit tradeoffs in transportation infrastructure expansion, specifically between spatial and social forms of connectivity, and between regional and local scale. We ground this framework in the case study of *Miaoqianzhijie*, a historic street in Guangzhou, China, which is planned to be demolished to construct a new subway interchange station. With empirical evidence, we argue that the expansion of subway infrastructure in urban areas produces tangible local social costs together with diffuse spatial benefits, constituting a network tradeoff that existing planning models inadequately capture. We operationalize the framework's social connectivity component through a participatory GIS survey that quantifies and visualizes how the historic street supports locals' place attachment and social ties. For the spatial connectivity component, we use an origin–destination flow analysis to illustrate how the spatial benefits of a new interchange station are distributed across the network and how modest the gains are for those living closest to the demolition site.

Specifically, we ask two research questions in the case study, 1) how can participatory GIS and social network mapping reveal the specific social infrastructure and ties jeopardized by demolition in a historic street? and 2) how are the spatial transit benefits of new subway expansions distributed across network users, and what tradeoffs arise for local versus regional connectivity in planning decisions?

This study contributes in three ways. Conceptually, it advances the network tradeoff framework by treating subway lines as spatial networks and historic streets as social networks in the same case, making tradeoffs between mobility gains and social connectivity losses more explicit. Methodologically, it demonstrates how participatory GIS-based social network mapping and OD-based network analysis can be combined to represent these tradeoffs. Practically and normatively, the study has an activist orientation: we elevated the voices of pro-preservation advocates through participatory data collection, visualization, and dissemination. Some findings have already been shared with the public and submitted to the local planning authority to inform decision-making. In doing so, we seek to provide language and tools that can inform more equitable infrastructure decisions, particularly in contexts where historic streets and everyday social networks are at risk.

In the following literature review, we situate this case study within broader debates on transportation infrastructure and social costs. We contrast place-based and spatial network framework, discuss their limitations, and argue for a network tradeoff framework that better reflects the intertwined nature of spatial and social connectivity in infrastructure planning.

## 2 Literature Review and Conceptual Framework

### 2.1 Place-based Approaches to Social Costs: Place Attachment, Social Infrastructure, Social Connectivity, Social Capital, and Gentrification

Transportation infrastructure expansion can reshape not only the physical environments, but also the social and economic ecosystems of nearby neighborhoods. Place-based approaches start from these lived consequences conceptualize "social costs" of infrastructure expansion in three broad types: (1) physical displacement, (2) environmental and quality-of-life impacts, and (3) community disruption (Meyer, 2016: 139–146). Physical displacement refers to the removal or relocation of residents and businesses. Environmental and quality-of-life impacts capture changes in air, noise, and water conditions as they intersect with everyday comfort and health. These two factors are typically measured by the number of displaced people or households and the scale of built environment changes. These measures align with regulatory requirements in environmental impact assessment, which specify reporting on relocation, land use change, noise, and pollution.

Unlike displacement counts or pollution levels, community disruption is harder to define and standardize. To understand what is lost when a historic street is demolished, it is useful to distinguish place attachment, social infrastructure, social connectivity, social capital, and gentrification as they are conceptualized and measured differently to capture community disruption.

Place attachment refers to people's functional and emotional bonds to specific locations, such as reliance on businesses, amenities, and services, but also a sense of belonging, identity, and collective memory (Tuan, 1977; Relph, 1976; Gurney et al., 2017). Social infrastructure denotes "third places", such as libraries, parks, small businesses, cafés, that foster interaction and sustain local identities and social ties (Oldenburg, 1989; Latham & Layton, 2019). Social infrastructure is the primary vehicle through which place attachment is sustained, and independent small businesses rely on localized customer networks and are particularly vulnerable to displacement (Shuman, 2001). Under this logic, social costs can be approximated by tallying the social infrastructures displaced and checking for replacement options.

In cultural mapping tradition, social connectivity describes the structure and density of relationships among people or between people and places, including how many ties they have, how often they interact, and where these encounters take place (Mehta, 2019). In transportation literature, classic street-level studies like Appleyard's Livable Streets show how higher traffic volumes drastically reduce unneighborly contacts and acquaintances, operationalizing social connectivity as counts and types of social ties on streets (Appleyard, 1981). Emerging urban big data enables more precise recording of social connectivity as it intersects or attaches to physical infrastructure. Studies have used Twitter follower network to measure barrier effects of highways (Aiello et al., 2025) and Yelp reviews to quantify how social relationships are attached to social infrastructure (Bendeck & Andris, 2022).

Social capital emphasizes the qualities and resources embedded in those relationships, such as trust, reciprocity, and access to support or information (Putnam, 2000), measured through survey scales of attachment or trust (Mak et al., 2023). Gentrification, by contrast, describes a process of socio-economic restructuring (i.e., rising property values, changing tenure, and shifts in class and demographic composition) that can weaken place attachments, disrupt existing social connections, and redistribute social capital even when residents are not formally displaced (Zuk et al., 2018; Dawkins & Moeckel, 2016).

Among these concepts, social connectivity most directly aligns with a network perspective, yet it remains under-measured and rarely visualized in studies of transport expansion. Preservation research seldom traces how specific projects reshape the social networks anchored in historic streets, while transport impact assessments rarely operationalize connectivity in place-based contexts. Existing data sources are also limited: ethnographic work captures only observable interactions, and social media data remain too sparse for site-level evaluation. This study therefore foregrounds social connectivity as the central mechanism of social cost which is interpreted alongside place attachment, social infrastructure, social capital, and gentrification, by documenting a snapshot of street-level social ties and juxtaposing them with the networked mobility impacts of the proposed subway station.

### 2.3 Spatial Network Approaches to Social Costs: System Engineering, Efficiency, and Coverage

Spatial network approaches adopt a systems-engineering logic, focusing less on specific places and more on the performance of the wider mobility network (Boeing, 2017; Guihaire & Hao, 2008; Akhavan et al., 2020). Such approach shifts the focus from localized disruptions to understanding infrastructure within a "space of flows" (Castells, 1996; Batty, 2013), symbolized as a system of nodes and edges facilitating the movement of people, goods, and information.

In the context of metro planning, subway lines and stations are modeled as a spatial network whose topology can be described with metrics such as degree, betweenness and closeness centrality, average shortest path length, network diameter, and global efficiency (Kim and Song, 2015; Feng et al., 2017; Derrible, 2012). The expansion of a subway system is therefore conceptualized as adding or modifying nodes and edges in this graph, and a proposal is judged by how it alters the connectivity, efficiency, and robustness of the entire network rather than by what happens on any single street.

Planning and transportation scholarship uses this spatial network perspective to define broad impact areas, affected populations, and measurable "costs" or objectives. In most complex-network analyses of subway systems, the impact area is the entire subway network, and the affected population is all riders whose origin–destination travel patterns might change when new stations or lines are added (Guihaire & Hao, 2008; Akhavan et al., 2020). Empirically, studies examine network-level impacts such as changes in average path length, diameter, and global efficiency, interpreted as increased or reduced travel times and detours for the system as a whole, and node-level impacts such as shifts in station centrality and accessibility, which indicate how flows and transfer pressures are redistributed across lines and hubs (Derrible & Kennedy, 2010; Derrible, 2012; Boeing, 2017). A related strand models vulnerability and resilience: here, the “social cost” of a design is operationalized as the loss of connectivity or accessibility under simulated failures or attacks on highly central stations and links, with more redundant networks viewed as more robust to disruption (Taylor, 2017; Cats & Jenelius, 2018).

Building on this, more recent work layers accessibility and equity onto the spatial-network view. Rather than evaluating metro expansions on travel time alone, studies incorporate monetary costs (fares), job accessibility, and socio-demographic differences to examine how rail investments redistribute accessibility benefits across neighborhoods and population groups (Costa et al., 2022; Yu & Cui, 2023). In these applications, subway stations are evaluated as nodes that extend coverage, shorten paths, relieve congestion at overloaded interchanges, or improve access for underserved areas.

However, this spatial network tradition often abstracts infrastructure as neutral nodes and edges, ignoring the differentiated physical and social context of specific locations. While this perspective is powerful for optimizing social costs in aggregated terms, such as system efficiency and resilience, they typically do not account for the fine-grained social infrastructures and connectivity that may be disrupted at specific sites where new infrastructure is built or existing streets are demolished. Even equity-oriented work tends to compare accessibility gains and losses across zones and demographic groups without representing how the same infrastructure that improves regional accessibility can simultaneously dismantle the street-level social networks that historic preservation and place-based research seek to protect.

### 2.4 The Network Tradeoff Framework

Bringing place-based and spatial network perspectives together reveals a core tension: the same infrastructure can simultaneously act as a connector for some flows and a barrier for others. This phenomenon has been described as network duality, which argues that network infrastructure often facilitates certain flows while excluding or degrading others (Andris et al., 2018; Liang, 2023).

In practice, such duality manifests as network tradeoffs between urban systems that support different types of connectivity at different speeds and different geographic scales. One form of the tradeoff arises between high-speed, regional mobility infrastructure (e.g., subways, highways, airports) and multimodal, local flows, such as walking, biking, or informal transit, due to incompatible built environment priorities (van Eldijk et al., 2020; van Eldijk et al., 2022; Orrick & Frick, 2013; Liang & Yang, 2026). A second form of tradeoff concerns transportation versus social infrastructure. Subway construction and associated redevelopment can displace existing social infrastructure, eroding local sociability and cultural life (Das, 2014; Lung-Amam et al., 2019). Because social networks are spatially embedded (Tao et al., 2022), infrastructure projects that displace community-supporting spaces can disrupt both local and more distant social connections, especially for low-income and socially disadvantaged groups whose ties are often close-knit and geographically concentrated (van Eijk, 2010). At the same time, subway stops and station-area public spaces can themselves function as social infrastructure, acting as hubs where diverse users encounter one another, linger, and participate in public life (Sadana, 2021; Latham & Layton, 2019). Third, a growing body of work focuses on the barrier effects of transport infrastructure on social networks: for example, one study uses Twitter follower networks to show that highways inhibit social ties, with a “barrier score” comparing observed versus simulated cross-highway ties and revealing strong disruptions, especially in historically marginalized Black neighborhoods (Aiello et al., 2025). Together, these examples highlight how the same project can simultaneously enable and dismantle connectivity across different networks.

Despite these studies, existing research and planning tools still tend to treat these dynamics one network at a time. Place-based studies document how historic streets and neighborhoods support social life but rarely formalize those relationships as networks that can be compared with regional mobility systems. Spatial network analyses of metro systems, in turn, rigorously model connectivity, efficiency, and even equity, yet largely treat stations and links as socially empty nodes and edges. Big-data studies of barrier effects hint at duality, but they typically focus on large linear infrastructures or aggregate neighborhood change rather than the fine-grained social networks anchored in a specific historic street. As a result, there is no common framework for representing both spatial and social connectivity within the same case and explicitly comparing gains in one against losses in the other. In contemporary planning practice, a network tradeoff framework is needed to bring these strands together, making visible how projects like a new subway station can simultaneously produce regional mobility benefits and concentrated local social losses, and to provide a language for debating whether, for whom, and under what conditions such tradeoffs are acceptable, which may be obscured in conventional impact assessments.

Motivated by this gap, we define the network tradeoff framework as a conceptual lens with three components: (1) representing both spatial and social connectivity as networks, where nodes and edges can be physical (stations, links) or social (people, relationships, social infrastructure); (2) juxtaposing gains and losses across these networks, interpreting infrastructure projects as simultaneously enabling and disrupting connectivity at different scales; and (3) motivating these patterns to impact planning choices, identifying missing data and uplift narratives for overlooked connectivity.

Under the network tradeoff framework, we conceptualize the social network as the web of relationships among people and places that converge on *Miaoqianzhijie*. Nodes are the small businesses and third places along the street, together with the people who own, work in, or regularly visit them; edges are the repeated visits, transactions, and social ties that these places host. We conceptualize the mobility network as the Guangzhou subway system, represented as a graph of stations (nodes) connected by rail links (edges) and used by origin–destination passenger flows. The network tradeoff framework therefore treats the *Miaoqianzhijie* project as a simultaneous reconfiguration of a street-level social network and a regional mobility network with concentrated losses in social connectivity along the historic street through the demolition or of high-attachment, low-replaceability social infrastructures, which are weighed against more diffuse gains in mobility connectivity for metro passengers who benefit from the new station. Grounding the framework in this case study allows us to examine both its contributions and its frictions with existing planning models.

This framework does not claim novelty in recognizing that both spatial and social networks matter. Rather, its contribution lies in bringing them into a single, network-based comparison. How the network tradeoff framework conceptualizes social impacts can have profound implications for how planners define affected populations, evaluate costs and benefits, and propose interventions (see Table 1 for comparison).

**Table 1.** Conceptual representation of the social costs of demolishing a historic street for a new subway station (case study) under three frameworks, which shows how this paper differentiates and complements the place-based and spatial network-based (or system engineering) perspectives.

| | **Place-based Framework** | **Spatial Network-based Framework** | **Network Tradeoff Framework** |
|---|---|---|---|
| **Theoretical Foundation** | Place attachment; Social infrastructure | Network science theory | Network duality |
| **Impact Area (What/Where is affected)** | The demolition planning boundary and surrounding neighborhoods (e.g., within a 1 km radius of the project site). | The entire subway network, including all stations connected to the proposed subway station. | Social or mobility connections in either the social or mobility network that may are rewired or severed, including the historic street segment and its web of social infrastructures and the people who regularly use them and the subset of the subway network whose accessibility and flows change with the new station. |
| **Population (who is affected)** | Residents within the defined impact area. | Populations whose origin-destination travel patterns would be affected by the proposed subway station. | Population for whom these specific nodes (street / station) are critical in their social or mobility networks, including those who maintain functional, social, or symbolic ties to the historic street and those whose travel patterns change because of the new station. |
| **Empirical Measurement of Social Impact (what is the cost; what to optimize)** | Number of displaced people/households<br><br>Land use and built environment changes<br><br>Loss of critical amenities affecting functional place attachment<br><br>Gentrification trends (shifts in property value and demographics) | Network-level: Inefficiency in mobility flows, increased travel time, and distance; Uneven node importance, leading to less resilient networks; Increased network diameter, reducing overall accessibility.<br><br>Node-level: Changes in node centrality and accessibility; Impact on travel efficiency to and from specific nodes. | Compare concentrated losses in the street's social network (removal of high-value social infrastructure and ties) against diffuse gains in regional mobility connectivity (who benefits from travel-time savings and improved connectivity). |
| **Equity Consideration** | Which demographic groups are most impacted by the demolition? | How to optimize network system performance by balancing accessibility with new subway stations and flow efficiency and resilience with new interchange stations through network design? | What types of connectivity (e.g., spatial or social) are prioritized in urban development? How can the regional travel benefits of new stations be balanced with the costs to local social connections? |
| **Challenges / Limitations** | Focus on impacts in the nearby environments | Focus on optimization within a single network and to the overall system performance rather than individual nodes. | Focus on bridging two differently conceptualized networks with distinct metrics, data sources, and visualization conventions; Lack of simple scalar metric to collapse social and mobility impacts into a single objective function; The transient nature of social connectivity that can be easily affected by shifting human dynamics and gentrification |
| **Policy and Action Implications** | Property right and compensation.<br><br>Relocation assistance | Plan new subway stations and lines strategically to account for both the network system and equity impacts, focusing on alleviating congestion, | Make dual-network impacts explicit in planning and support planners, residents, and decision-makers in debating not only whether a project improves accessibility, but whose |
| | Inclusive land uses and new amenities<br><br>Historical or cultural preservation protection | increasing alternative routes (reducing dependency on individual transfer stations) for resilience, and extending network coverage to underserved areas. | mobility is enhanced, whose social connectivity is degraded, and how alternative designs could rebalance that tradeoff.<br><br>Articulate and elevate the value of social connectivity upheld by the historic street; Acknowledge community sacrifices and negotiate reparative actions (e.g., oral history and memory projects) that respond to the specific social networks disrupted, not just to the number of units or parcels lost.<br><br>Design to preserve social connectivity where possible, such as preserving small businesses that are important to local social life or relocate social infrastructure nearby. |

## 3 Case Study: *Miaoqianzhijie* and *Shuqianlu* Subway Station

*Miaoqianzhijie* (jie means street in Chinese) is located in downtown Guangzhou, China (population 19 million). Situated in the *Xinhepu Cultural District*, the street is narrow and walkable, about 200 meters long and 9 meters wide (Figure 1), and lies within walking distance of four schools, a Children's Palace (for youth activities), and two subway stations (*Dongshankou* and *Donghu*). It hosts a mix of residential, commercial, and institutional uses. The first store opened in 1915, with more appearing as overseas Chinese returned to Guangzhou between the 1920s and 1960s. *Memory Dongshan* (2020), an oral history book, features multiple interviews with elderly residents recalling their experiences with the street's stores.

In September 2020, the Guangzhou government released a draft plan to demolish all buildings on the north side of *Miaoqianzhijie* and an adjacent road to construct *Shuqianlu* Station on Line 10. Line 10 is a new subway line, under construction since 2018 and partially operational since 2025, designed to connect the city's southwest and northeast while offering alternative routes to the financial center. The plan connects *Shuqianlu* with *Dongshankou* Station, creating an interchange for Lines 1, 6, and 10. Located near the center of the line, *Shuqianlu* Station is intended to relieve congestion from overloaded transfer stations (Figure 1). The proposed construction would tunnel beneath *Miaoqianzhijie* and demolish the north side of the street to accommodate the interchange, displacing all small businesses in the affected area. At the time of the draft plan, no alternative design was presented to the public.

The draft sparked public pushback, with citizens questioning the station's necessity and calling to preserve *Miaoqianzhijie* as a site of collective memory. In response, Guangzhou Subway Company defended the plan on social media, citing studies that the design "improves coverage and efficiency" and that the station is reasonably spaced from adjacent ones (1.6km and 1.3km). They also noted the site lies outside Xinhepu Cultural District's core protection zone and that no buildings qualify for historic preservation.

In November 2020, volunteers, including authors of this paper, formed the *Miaoqianzhijie Research Group* to document and share the street's social value. Little public information was available on planning histories, compensation, or relocation. The fate of tenants and business owners, many of whom are not property owners, remains uncertain.

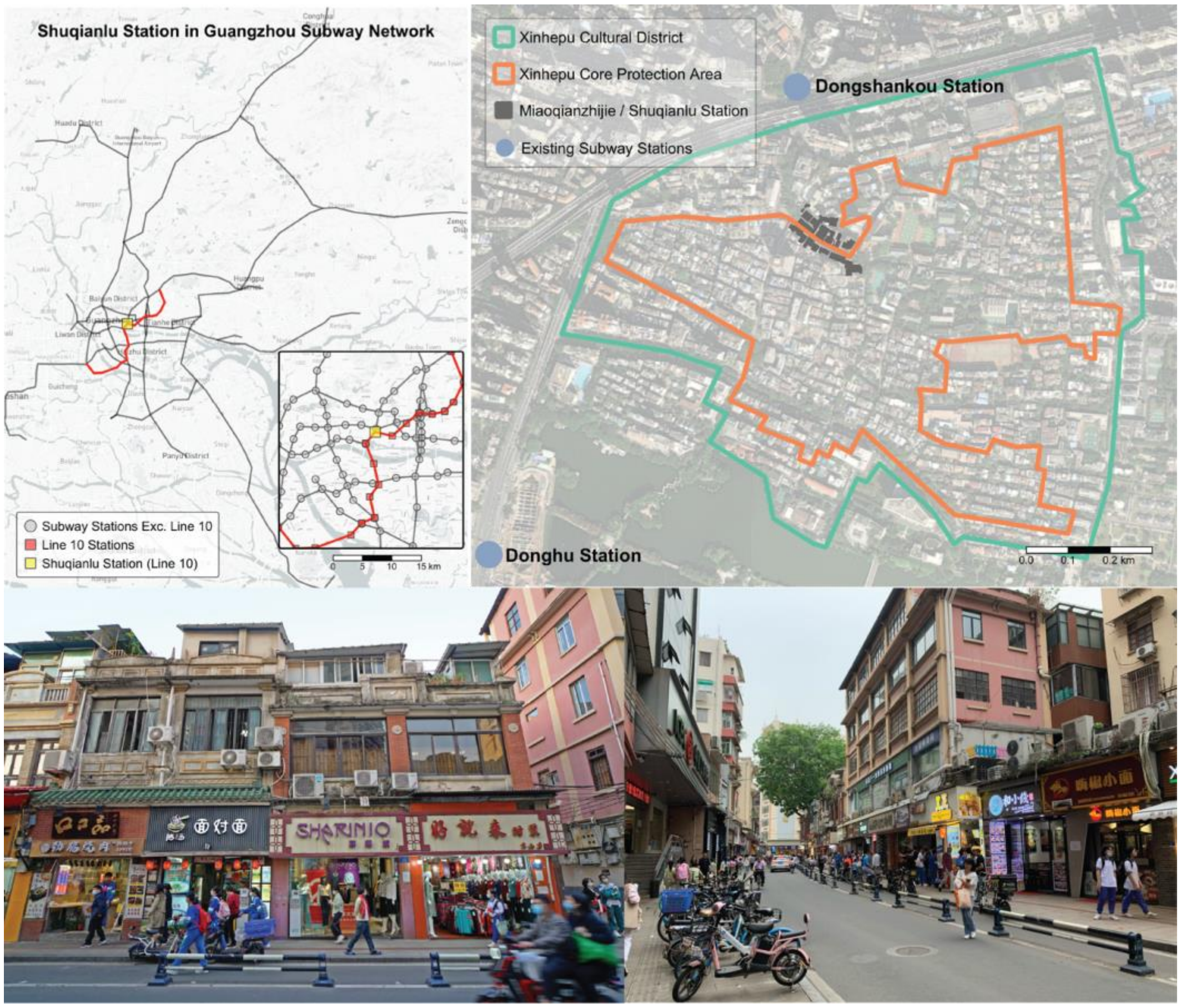

**Figure 1.** (from top left to bottom right) The location of *Shuqianlu* Station (proposed station) in the Guangzhou subway network, the location of *Miaoqianzhijie* (*Shuqianlu* Station) and nearby subway stations in *Xinhepu Cultural District*, and the north streetscapes of *Miaoqianzhijie.* Photo credit: *Miaoqianzhijie Research Group* volunteers, 2021.

## 4 Data and Methods

### *4.1* Understanding the Context: Geospatial and Statistical Data of *Miaoqianzhijie*

There is no official, open data about *Miaoqianzhijie* or small businesses located along the street. To collect these data, we conducted two field trips in February 2021 and created a database for 82 small businesses on *Miaoqianzhijie*, including their storefront pictures, addresses, store names, building levels, business types, and year opened (if available). The geospatial shapefiles of the stores and residential buildings were drawn based on satellite images of the area and validated against field volunteers' context knowledge. These data shape our understanding of the study area and provide a base layer for survey respondents to identify their interactions with *Miaoqianzhijie*.

### 4.2 Measuring the Social Network Cost: A Participatory GIS Survey and Informal Interviews

Under the network tradeoff framework, we measure the social cost of demolition as the removal or weakening of nodes and ties in *Miaoqianzhijie's* street-level social network. Small businesses along the street are treated as social infrastructure nodes and owners, workers, nearby residents, and frequent visitors are the people connected to them. Edges are the repeated visits, transactions, and social relationships these places host.

To operationalize this, we designed a spatially embedded GIS survey on the Maptionnaire platform (Figure 2), supplemented by informal interviews. Unlike conventional surveys that ask respondents to rate their social interactions or sense of place in the abstract, our interface required respondents to pin their social network onto the street itself. Each response captures spatial, relational, and qualitative information about a single edge, linking people to places and people to each other, rather than only recording aggregate attitudes. Taken together, the survey allows us to approximate the social network cost of the project not simply as "number of businesses lost," but as the targeted removal of especially central, high-attachment, low-replaceability social infrastructures and the dense webs of relationships they sustain.

The survey has three sections. The first section collects respondents' age, their relationships to the street (e.g., property owners, frequent visitors, etc.), and their perceived importance of the street to their identity, personal growth, daily life functions, social ties, and Guangzhou's social and cultural history. These questions provide node- and person-level attributes that indicate overall place attachment and who is most dependent on the street's social infrastructure.

The second section allows the respondents to pin places on the maps to highlight their visits and comment on the trip purpose, trip companion (i.e., who they visit the store with), and the replacement for the visit if the place is demolished. Respondents can also pin their memories on the street. This information is used to map edges in the social network: which businesses are linked to which users, for what activities, and in whose company. It also yields a replaceability measure for each establishment, signaling which nodes are high-attachment, low-substitute anchors in the network. For respondents who identify as direct stakeholders of *Miaoqianzhijie* (property owners, residents, business owners, workers), we additionally ask which stores they interact with and how often they help or receive help from others on the street. These responses allow us to derive a small-scale social network within *Miaoqianzhijie*, highlighting patterns of mutual aid, familiarity, and everyday cooperation. The last section is open-ended, asking respondents to discuss how the demolition may impact their lives.

We distributed the participatory GIS survey twice, each with one month, from April to May 2021 and September to October 2021. The digital version is accessible to the public through mobile or web interfaces. We also printed paper versions and distributed them in-person at *Miaoqianzhijie* to passengers and store owners. We advertised the survey through social media, personal networks, and posters at a local bookstore and a café shop and rewarded respondents with custom-made postcards. All surveys are written in Mandarin.

In total, we received 200 valid responses, of which 35 copies were transcribed from paper-based surveys and the rest from the web. Among people who provided identity (n=151), 17% say they are direct stakeholders of *Miaoqianzhijie* and 58% are frequent visitors; we refer to this population as locals when reporting survey statistics. Among people who provided age (n=177), 53% of the respondents are between 18-25 years old, 41% are between 26-50 years old, and only 6% are elderly beyond 50 years old. To complement the survey responses, we also conducted informal interviews with store owners and six informants to understand how they experience *Miaoqianzhijie* daily and their perceptions of demolition's social impact.

It is worth noting that China was experiencing severe COVID-19 shocks during the time of data collection (2021), which limits the scale, quality, and demographic representations of the survey responses we received. Due to travel restrictions, we trained 12 volunteers at *Miaoqianzhijie Research Group* to conduct much of the fieldwork, informal interviews, and survey distribution.

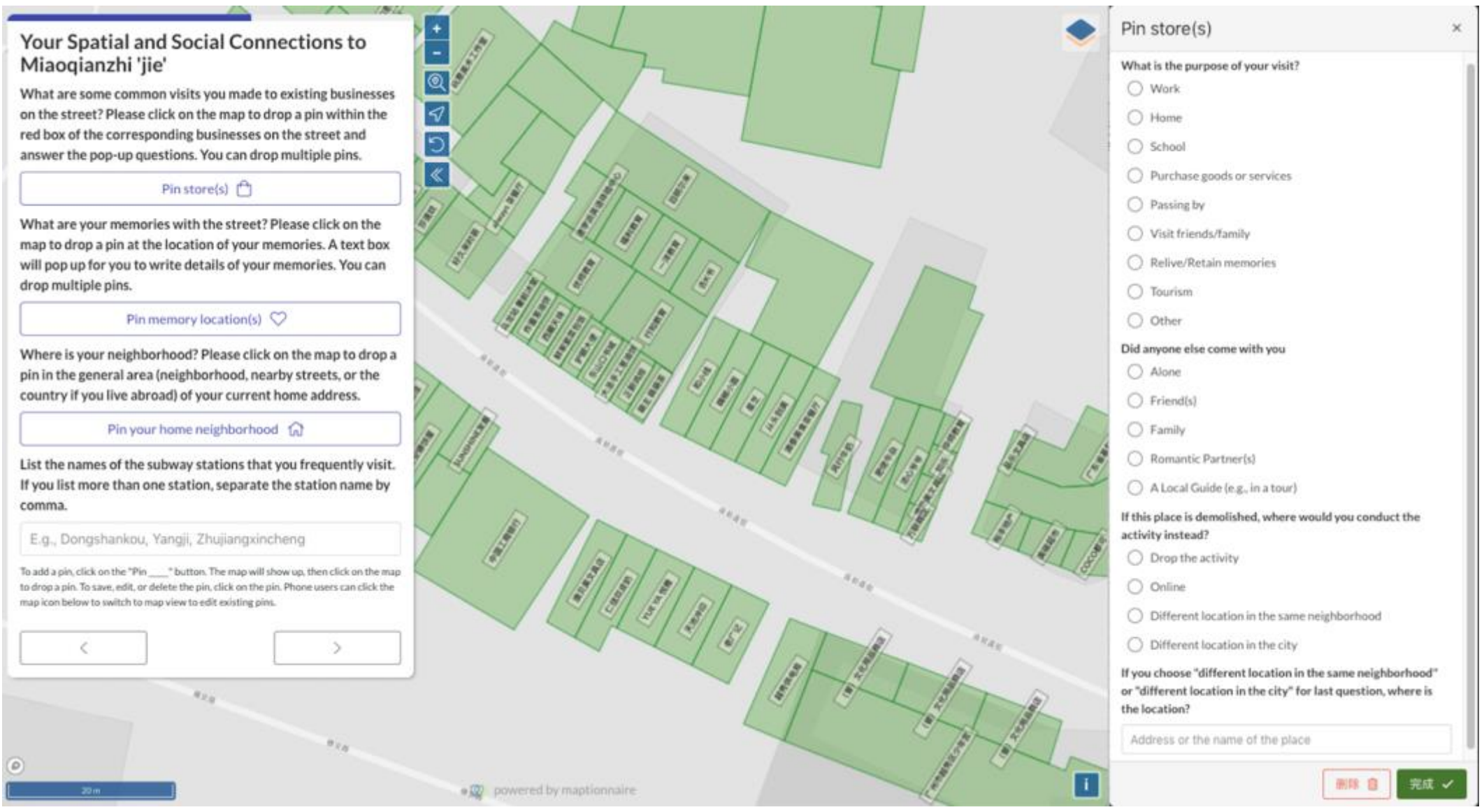

**Figure 2.** Interface of the participatory GIS survey and pop-up questions when respondents pin a store (green polygon) on the map.

### 4.3 Comparing the Spatial Benefits of a New Subway Station: Origin-Destination Flow Analysis

We define the spatial benefits of the new subway station as reduced travel time and increased route options for locals to access their destinations. We collected origin and destination (OD) flow data of the Guangzhou subway system from a published paper (Ye et al., 2021). This raw data was obtained from Guangzhou Subway Company's smart card system, which records the entry and exit stations for each trip. The aggregated data includes the number of trips for each pair of subway stations on a weekday in April 2018.

We use OD data from *Dongshankou Station* (five minutes walking distance from *Miaoqianzhijie*) to represent the locals' existing travel demands. We assume that the travel demands stay constant with the new Line 10 and thus calculate two metrics for each subway station to measure the OD difference to compare mobility benefits in subway network scenarios with and without Line 10. The first metric is the percentage of trips that can reduce more than one hop to reach destinations, and the second metric is the percentage of trips that have increased route options to their destinations with an equal or smaller number of hops. A hop refers to travel from one station to its adjacent station in the subway network. The costs of transferring between lines are neglected. These metrics capture the spatial benefits of *Shuqianlu Station* (and Line 10) through reduced travel distance and increased travel options, respectively. Finally, we compared these two metrics across stations to determine whether locals whose trips represented by *Dongshankou Station* will receive more spatial benefits than passengers from other stations.

## 5 Results

### 5.1 The Social Impact of Demolition: Loss of Social Connectivity

To understand the social impact of demolition, we examine how *Miaoqianzhijie* functions as a corridor of social infrastructure. Documentation of 89 small businesses reveals a diverse landscape of services tailored to nearby residents and visitors (Figure 3), including education and cultural services (n=28, 31.5%), snacks and drinks (n=18, 20.2%), life essentials (n=13, 14.6%), sit-down restaurants (n=12, 13.5%), and clothing and shoes (n=9, 10.1%). The north side of *Miaoqianzhijie* (location for demolition and home to 57 businesses) has a particularly high concentration of education and light dining services, serving as a transitional space for thousands of students, parents, and teachers from nearby institutions.

The composition of businesses further reflects the local-serving nature of the street. In contrast, Beijing Road, another historic pedestrian commercial street in Guangzhou, has 47.2% of its points of interest (POIs) concentrated in shopping and consumption categories, including shopping malls, beauty and clothing stores, electronic goods, and cultural retail (Xie et al., 2024). This distribution contrasts with the education- and food-oriented POI structure observed in *Miaoqianzhijie*. Most of *Miaoqianzhijie*'s businesses are small, independently owned, and provide affordable, customized services for elderly residents and students. For example, six businesses have operated for over ten years, and many offer long-standing services tailored to the local community. In network terms, these small businesses define the node set through which everyday life and social ties are routed, which are not easily substitutable “units” but high-attachment social infrastructure nodes in the street’s social network.

Survey results further show how residents and visitors attach their routines, encounters, and social relationships to this node set. Sixty-two percent of local respondents (vs. 52% of all respondents) said *Miaoqianzhijie* is important to their daily activities and 55.7% of local respondents (vs. 45.4% of all) said the street supports their social ties. Perceived importance of *Miaoqianzhijie* for personal growth, daily life, and social ties increases with age, local identity, and duration of engagement. In contrast, cultural importance is rated higher by younger, non-local respondents and those with shorter (1–3 year) relationships. On one hand, this reversal likely reflects sampling bias, as non-local respondents who are often recruited through personal networks or social media, may express generalized cultural appreciation despite limited lived experience with the street. On the other, it suggests that *Miaoqianzhijie* supports multi-scalar social connectivity: it enables intimate local interactions while also contributing to broader place identity.

**Table 2:** Distribution of Miaoqianzhijie’s importance by demographic groups, identity, and length of relationship

| Category | n | Importance: Personal | Importance: Life | Importance: Social | Importance: GZ Culture | Pct > = 3 (Personal) | Pct > = 3 (Life) | Pct > = 3 (Social) | Pct > = 3 (GZ Culture) |
|---|---|---|---|---|---|---|---|---|---|
| Age: 18-25 | 108 | 2.74 (1.20) | 2.38 (1.14) | 2.26 (1.18) | 3.24 (0.85) | 65.88% | 52.94% | 44.71% | 84.88% |
| Age: 26-50 | 89 | 2.59 (1.17) | 2.37 (1.11) | 2.28 (1.08) | 3.17 (0.91) | 57.14% | 49.30% | 45.07% | 78.87% |
| Age: 51+ | 18 | 2.65 (1.00) | 2.59 (1.00) | 2.28 (1.27) | 3.17 (1.04) | 64.71% | 58.82% | 50.00% | 77.78% |
| Identity: Local | 154 | 2.92 (1.07) | 2.62 (1.04) | 2.51 (1.08) | 3.17 (0.86) | 72.66% | 61.70% | 55.71% | 81.12% |
| Identity: Non-local | 58 | 2.09 (1.26) | 1.76 (1.00) | 1.76 (1.04) | 3.33 (0.85) | 36.96% | 24.44% | 21.74% | 86.67% |
| Length of Relationship (Local Only): < 1 yr | 32 | 2.36 (1.08) | 2.26 (0.90) | 2.08 (0.89) | 2.71 (0.66) | 44.00% | 51.85% | 38.46% | 67.86% |
| Length of Relationship (Local Only): 1-3 yrs | 42 | 2.59 (1.21) | 2.28 (1.07) | 2.40 (1.03) | 3.40 (0.67) | 61.54% | 48.72% | 47.50% | 90.00% |
| Length of Relationship (Local Only): 4-10 yrs | 29 | 3.12 (0.88) | 2.52 (1.12) | 2.36 (1.22) | 3.28 (0.84) | 84.00% | 52.00% | 56.00% | 84.00% |
| Length of Relationship (Local Only): 10+ yrs | 51 | 3.36 (0.83) | 3.12 (0.85) | 2.90 (1.03) | 3.18 (1.00) | 90.00% | 82.00% | 71.43% | 80.00% |
| All | 215 | 2.67 (1.17) | 2.39 (1.11) | 2.27 (1.14) | 3.21 (0.89) | 62.21% | 52.02% | 45.40% | 81.71% |

Notes: The column reads as the number of valid responses, the average and standard deviation of importance rating [0, 4] of Miaoqianzhijie to the respondent's identity and personal growth, daily life functions, social relationships, and Guangzhou's social and cultural history, and the percentage of people that believes that Miaoqianzhijie is somewhat important (value = 3) or very important (value = 4) to these dimensions. The row reads as age groups, how the respondent identifies with Miaoqianzhijie, including being a property owner, business owner, staff, residential renters, (nearby) residents, and frequent visitor (all of which classified as local), or being a general Guangzhou resident, oversea Chinese immigrant, short-term visitor/tourist, and others (all of which classified as non-local). Distribution statistics and percentages are calculated for each category by row. Category totals may not sum to the full sample size because sociodemographic information was optional in the survey.

Using survey data from 278 store visits and 131 location memories, we examine how visits are structured by companionship and how these place-based relationships might be redistributed if the street’s social infrastructure were demolished (Table 3). We also mapped how each building supports social relationships and the interactions of primary stakeholders who live and work at *Miaoqianzhijie* (see Figure 3), revealing the intensity of social connectivity.

Seventy-one percent of visits are made with others rather than alone, including with friends (42%), family (19%), and romantic partners (9%). These shared visits are concentrated in stores selling snacks and drinks, restaurants, and, to a lesser extent, education-related services. When asked how these connections might be reestablished if the stores were removed, 39% of visits would be dropped, 5% shifted online, and 29% relocated elsewhere in the city, while only 26% could be reproduced nearby. This pattern suggests limited substitutability and highlights the role of these establishments in supporting both daily routines and place-based attachments (see Table 3). For example, the most “irreplaceable” business was a 30-year-old Thai restaurant, while chains like milk tea stalls were seen as more easily replaced. This finding supports the notion that social value increases over time and with uniqueness and some businesses become high-attachment, low-substitute nodes in the street’s social network.

These attachments are unevenly distributed across stores. Only 59 of the 89 businesses were identified in the survey, and most visits are concentrated in snacks and drinks and restaurant categories, with lower engagement in other categories. Different types of relationships are also anchored in different store types. Interpreting visits by relationship (column-wise), visits with friends (92/118 ~ 78%) and romantic partners (18/25 ~ 72%) are concentrated in snacks and drinks and restaurants, whereas family visits are more evenly distributed, including notable shares in education (12/54 ~ 22%) and essential services (11/54 ~ 20%) (see Table 3). Figure 3 also revealed that some businesses also support all three relationships. In pinned memories, a parent recalled that she usually waits at the *Dongshankou bookstore* for her kid to go off school or asks her kid to wait there if she is late to pick up. Other students also mentioned going to the bookstore with friends to read for free before their night studies. These patterns show that some businesses are multi-relational nodes that concentrate different kinds of ties, such as family, friendship, and romantic companionship, while others play more specialized roles.

**Table 3:** Distribution of Miaoqianzhijie's visits by store categories, purpose, companionship, and where the respondent could rebuild this connection.

| Store category | N stores | N visited | N visits | With whom: Alone | With whom: Friend(s) | With whom: Family | With whom: Romantic partner(s) | Rebuild: drop activity | Rebuild: online | Rebuild: nearby | Rebuild: elsewhere in city |
|---|---|---|---|---|---|---|---|---|---|---|---|
| Snacks and Drinks | 18 | 18 | 108 | 32 (30%) | 49 (45%) | 18 (17%) | 9 (8%) | 37 (35%) | 4 (4%) | 40 (37%) | 26 (24%) |
| Restaurants | 12 | 11 | 79 | 15 (19%) | 43 (54%) | 11 (14%) | 9 (11%) | 41 (52%) | 2 (3%) | 13 (16%) | 23 (29%) |
| Education | 28 | 13 | 48 | 16 (33%) | 17 (35%) | 12 (25%) | 3 (6%) | 19 (40%) | 6 (12%) | 7 (15%) | 16 (33%) |
| Essentials | 13 | 8 | 32 | 12 (38%) | 6 (19%) | 11 (34%) | 3 (9%) | 7 (22%) | 1 (3%) | 11 (34%) | 13 (41%) |
| Clothing | 9 | 5 | 7 | 3 (43%) | 1 (14%) | 2 (29%) | 1 (14%) | 3 (43%) | 1 (14%) | 1 (14%) | 2 (29%) |
| Other | 9 | 4 | 4 | 2 (50%) | 2 (50%) | 0 (0%) | 0 (0%) | 2 (50%) | 1 (25%) | 0 (0%) | 1 (25%) |
| All | 89 | 59 | 278 | 80 (29%) | 118 (42%) | 54 (19%) | 25 (9%) | 109 (39%) | 15 (5%) | 72 (26%) | 81 (29%) |

Notes: Columns report the following: number of stores on the street in each category; number of stores identified as visited in the survey; total number of visits; number of visits made alone or with friends, family, or romantic partners; and how each people–store relationship could (or could not) be reestablished if the street is demolished (i.e., drop the activity, move online, relocate nearby in the same neighborhood, or relocate elsewhere in the city). Percentages are calculated for each store category by row.

To depict a person-to-person network layered on top of these person-to-place ties, we further visualized the social interactions between Miaoqianzhijie's primary stakeholders (i.e., those who own property and live or work there) in Figure 3, inspired by Appleyard's work (1981). Each (anonymized) line represents common interactions between respondents, such as shopping, visiting, helping, spanning across short and long distances along the street. The quality of these ties points to substantial social capital embedded in Miaoqianzhijie's social network. Sixty-four percent of respondents reported giving or receiving help from others on the street, indicating reciprocal, embedded interactions rather than purely transactional encounters. For instance, during fieldwork, a scallion pancake vendor posted a sign reminding customers to leave space for bookstore visitors next door, reflecting everyday mutual care and coordination. While this dataset is only a limited snapshot of the social network within *Miaoqianzhijie*, it already reveals dense ties crisscrossing the north and south sides of the street as a single social whole, challenging the argument that the north side can be cleanly separated from preservation considerations.

Pinned memories reveal that small businesses on *Miaoqianzhijie* also rely on acquaintances, local customers, and institutions nearby. For instance, a respondent recalled that he has known the juice store owner for a long time and the owner also remembers many students who went to schools nearby and would give them special discounts. A few business owners told us that they have fixed customers from neighboring communities for decades. These findings suggest that Miaoqianzhijie's social value lies in the structure of its network, such as interdependent nodes, repeated encounters, and mutual support, and demolition risks breaking the relational pathways that make everyday life function.

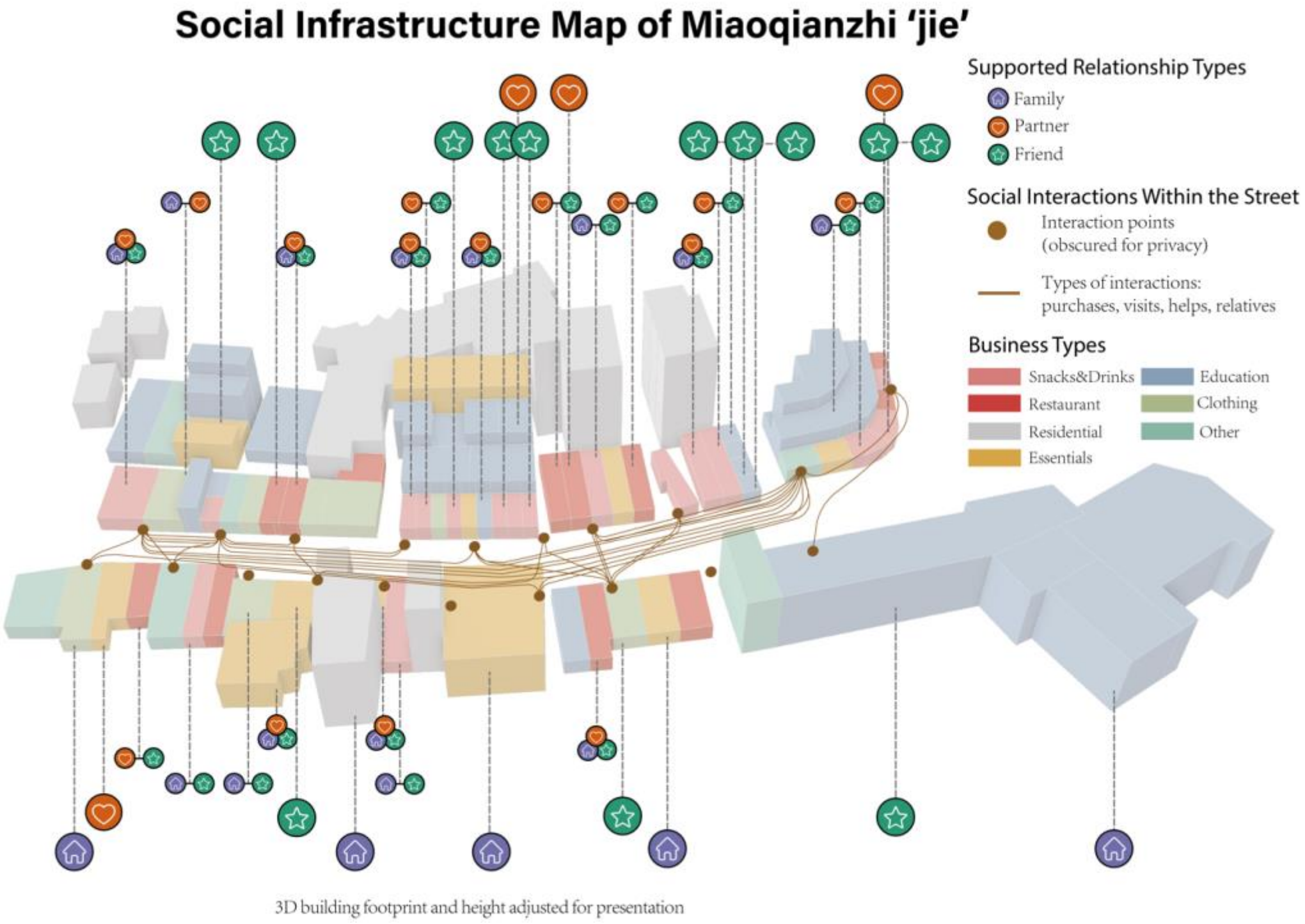

**Figure 3.** Social Infrastructure Map of *Miaoqianzhijie* (2021). Each 3D polygon represents a small business along *Miaoqianzhijie*, color-coded by type. Lines represent different forms of social interaction identified through surveys and informal interviews. To protect privacy, respondent locations are abstracted to evenly distributed interaction points spaced along the street. Symbols indicate the types of relationships supported by each business, based on whether respondents reported visiting these locations with family, partners, or friends.

### 5.2 The Spatial Benefits of the New Subway station: Gains in Mobility

To evaluate the spatial connectivity benefits of *Shuqianlu* Station and Line 10, we compared the reductions in travel distance and increased route options for station-to-station trips across the system (see Figure 4, Table 4). We used *Dongshankou* Station (200 meters from Shuqianlu Station) as the closest proxy to observe potential benefits to locals. Located in the city center, Dongshankou Station has high human traffic, with a total OD of 63,102, ranking the 73 percentiles of flow volume across all stations.

Figure 4 (left) indicates that passengers living along the north-south direction of Line 2 (blue), Line 6 (teal green), and Line 8 (teal) reduce travel distances the most through Line 10. Overall, Line 10 helps 1.3% of station pairs shorten their distances with more than one hop. Among all stations, *Zhongda* Station benefits the most: it reduces distances (>1 hop) to 30 stations (out of 195) and has the highest percentage of flows (21.4%) benefited (see Table 4). In contrast, *Dongshankou* station, which is the closest station to *Miaoqianzhijie* which represents the locals' current travel demand, reduces distances (>1 hop) to only 18 stations and benefits 5.3% (ranked 11) of flows. When considering only major distance reductions (>2 hops), the benefit for *Dongshankou* station drops to just 1%.

In terms of increased route flexibility (i.e., more than one alternative route to the same destination), the results are similarly modest (Figure 4, right). Different from travel distance, the benefits of increased route options (>1 route) are more distributed across stations. While 23% of station pairs benefit from Line 10's expanded routing options, *Dongshankou* station ranks only 26th among all stations, with 15.5% of flows gaining additional route options. Again, *Zhongda* Station benefits the most: 32.4% of flows originating from this station would have at least two more routes (>1 route) to reach their destinations with the same or fewer hops.

We also tested the spatial benefits of an alternative scenario for Line 10 in which *Dongshankou* Station is bypassed while the rest of the line remains intact. We expect that any trips that pass through *Dongshankou* Station will have distance reductions but lose route diversity since the direct interchange between Line 10 and Lines 1 and 6 is eliminated. This expectation is partially supported by the results, particularly for stations that are structurally dependent on *Dongshankou*. For example, the percentage of trips from *Tuanyida Guangchang* with distance benefits increases by 11.9 percentage points in the alternative scenario. A deeper analysis confirms this dependency: 53.60% of its origin flows have *Dongshankou* on at least one shortest path, and 12.26% of its flows experience a change in shortest-hop count between the two scenarios. In contrast, *Zhongda* exhibits lower dependency and almost no change in shortest-hop structure (0.18% of flows)

At the same time, the results reveal important nuances that deviate from this expectation. In particular, some stations experience increases in route options after *Dongshankou* is removed. This occurs because the counterfactual does not simply eliminate connectivity but reconfigures it by introducing bypass links between neighboring stations. As a result, flows that were previously funneled through *Dongshankou* may be redistributed onto more direct or alternative paths. This mechanism explains why stations such as *Wuyangcun* can create additional feasible routes even in the absence of the interchange, leading to an increase in the percentage of trips with multiple route options. Together, these findings suggest that while removing a major interchange generally reduces connectivity, its net effect depends on how the network is rewired, with some areas benefiting from the release of path dependencies and the creation of alternative connections.

In sum, the spatial connectivity gains from Line 10 are unevenly distributed, with *Dongshankou* Station and its surrounding community receiving relatively limited improvements compared to other areas. While the expansion enhances regional accessibility, its benefits to local users affected by demolition are modest. The alternative scenario further demonstrates that Line 10 can still deliver substantial spatial benefits across the network without relying on *Dongshankou* as an interchange. These results reinforce the core tension of this study: the social costs are concentrated and tangible, while the spatial benefits are diffuse and unevenly realized at the local level, which is a critical network tradeoff in infrastructure planning.

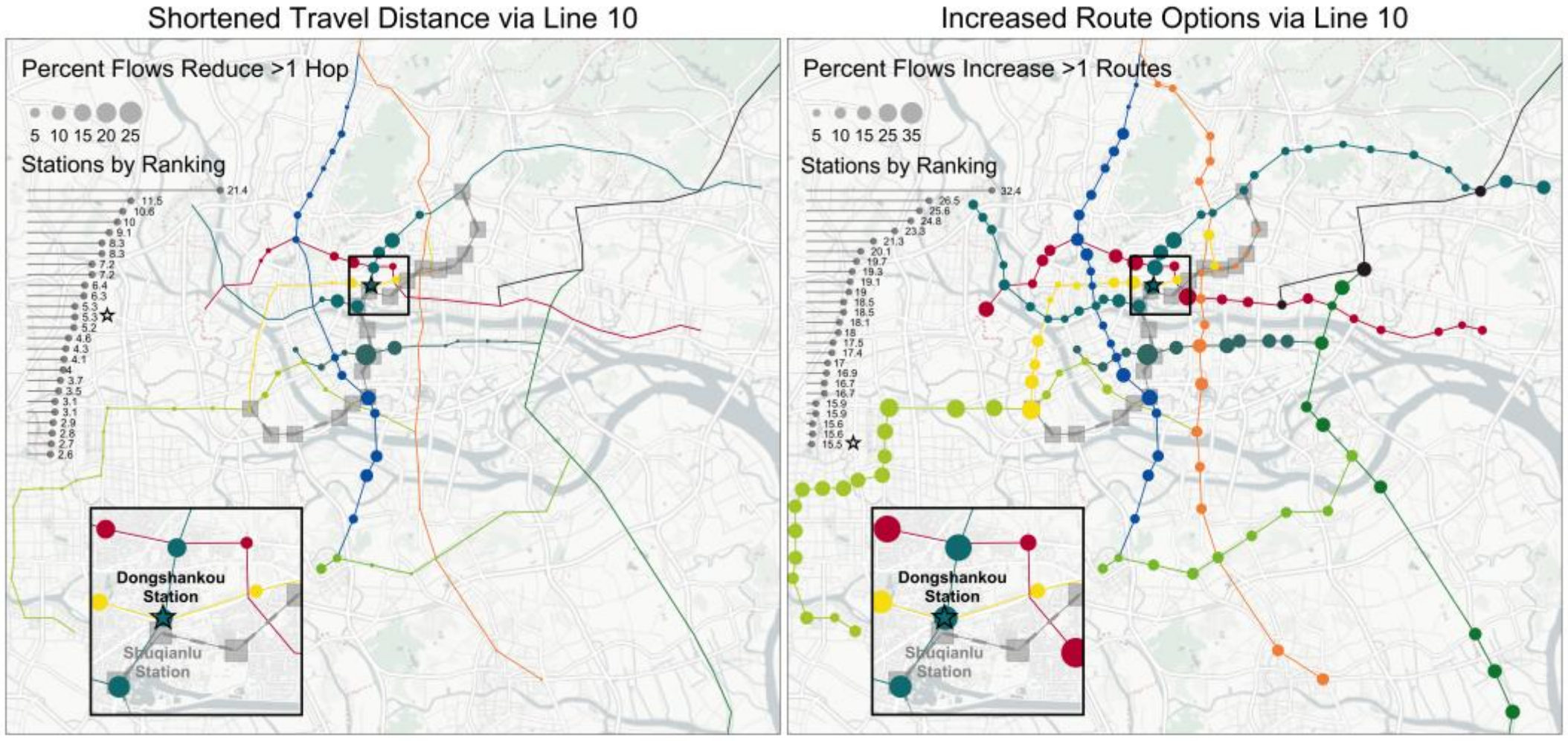

**Figure 4.** (left) Percentage of flows (trips) that have reduced more than one hop via Line 10; (right) Percentage of flows (trips) that has increased more than one route to reach the same destination without increasing network hops. Lollipop graphs show the ranking of the top 26 stations with the highest percentage of flows benefited via Line 10. *Dongshankou* Station is marked with a star.

**Table 4.** Top 26 stations ranked by the percentage of origin-only flows that benefit from the opening of Line 10, either through reduced travel distance (>1 hop) or increased route options (>1 route) to the same destinations. Columns also report total OD flows and the same metrics under an alternative scenario in which Dongshankou Station is removed from Line 10, along with the resulting differences. Statistics for Dongshankou Station and those referenced in the text are bolded.

| **Percentage of flows w/ distance reduced > 1 hop (Top 26)** | | | | **Percentage of flows w/ route increase > 1 route (Top 26)** | | | |
|---|---|---|---|---|---|---|---|
| Station | Total OD | Build Line 10 | Line 10 w/o New Interchange (Δ) | Station | Total OD | Build Line 10 | Line 10 w/o New Interchange (Δ) |
| **中大 (Zhongda)** | **79110** | **21.4%** | **21.5% (Δ +0.2 pp)** | **中大 (Zhongda)** | **79110** | **32.4%** | **37.0% (Δ +4.6 pp)** |
| 沙河顶 (Shaheding) | 47700 | 11.5% | 11.5% (Δ +0.0 pp) | 西塱 (Xilang) | 57866 | 26.5% | 26.5% (Δ +0.0 pp) |
| 东晓南 (Dongxiaonan) | 96139 | 10.6% | 10.8% (Δ +0.2 pp) | 金融高新区 (Jinrong Gaoxinqu) | 25180 | 25.6% | 25.6% (Δ +0.0 pp) |
| 鹭江 (Lujiang) | 72862 | 10.0% | 10.0% (Δ +0.0 pp) | 龙溪 (Longxi) | 21625 | 24.8% | 24.8% (Δ +0.0 pp) |
| **团一大广场 (Tuanyida Guangchang)** | **31203** | **9.1%** | **21.0% (Δ +11.9 pp)** | **五羊邨 (Wuyangcun)** | **85246** | **23.3%** | **28.7% (Δ +5.4 pp)** |
| 南浦 (Nanpu) | 56131 | 8.3% | 8.3% (Δ +0.0 pp) | 菊树 (Jushu) | 16778 | 21.3% | 21.3% (Δ +0.0 pp) |
| 东湖 (Donghu) | 34343 | 8.3% | 17.6% (Δ +9.4 pp) | 沙河顶 (Shaheding) | 47700 | 20.1% | 18.8% (Δ -1.3 pp) |
| 黄花岗 (Huanghuagang) | 23755 | 7.2% | 7.2% (Δ +0.0 pp) | 千灯湖 (Qiandenghu) | 29531 | 19.7% | 19.7% (Δ +0.0 pp) |
| 洛溪 (Luoxi) | 58149 | 7.2% | 7.2% (Δ +0.0 pp) | 淘金 (Taojin) | 65160 | 19.3% | 14.9% (Δ -4.4 pp) |
| 区庄 (Quzhuang) | 62139 | 6.4% | 6.4% (Δ +0.0 pp) | 朝安 (Chaoan) | 13087 | 19.1% | 19.1% (Δ +0.0 pp) |
| 淘金 (Taojin) | 65160 | 6.3% | 6.3% (Δ +0.0 pp) | 区庄 (Ouzhuang) | 62139 | 19.0% | 14.9% (Δ -4.1 pp) |
| 烈士陵园 (Lieshilingyuan) | 114660 | 5.3% | 5.3% (Δ +0.0 pp) | 南桂路 (Nanguilu) | 27129 | 18.5% | 18.5% (Δ +0.0 pp) |
| **东山口 (Dongshankou)** | **63102** | **5.3%** | **5.3% (Δ +0.0 pp)** | 东晓南 (Dongxiaonan) | 96139 | 18.5% | 19.9% (Δ +1.4 pp) |
| 广州南站 (Guangzhou Nanzhan) | 141313 | 5.2% | 5.2% (Δ +0.0 pp) | 滘口 (Jiaokou) | 82838 | 18.1% | 18.1% (Δ +0.0 pp) |
| 南洲 (Nanzhou) | 37595 | 4.6% | 4.6% (Δ +0.0 pp) | 赤岗 (Chigang) | 44082 | 18.0% | 20.3% (Δ +2.3 pp) |
| 会江 (Huijiang) | 26804 | 4.3% | 4.3% (Δ +0.0 pp) | 普君北路 (Pujun Beilu) | 17033 | 17.5% | 17.5% (Δ +0.0 pp) |
| 江泰路 (Jiangtailu) | 61446 | 4.1% | 4.4% (Δ +0.3 pp) | 黄花岗 (Huanghuagang) | 23755 | 17.4% | 14.5% (Δ -2.9 pp) |
| 小北 (Xiaobei) | 73861 | 4.0% | 4.0% (Δ +0.0 pp) | 祖庙 (Zumiao) | 38881 | 17.0% | 17.0% (Δ +0.0 pp) |
| 石壁 (Shibi) | 7210 | 3.7% | 3.7% (Δ +0.0 pp) | 桂城 (Guicheng) | 22054 | 16.9% | 16.9% (Δ +0.0 pp) |
| 天平架 (Tianpingjia) | 29251 | 3.5% | 3.5% (Δ +0.0 pp) | 季华园 (Jihuayuan) | 16506 | 16.7% | 16.7% (Δ +0.0 pp) |
| 昌岗 (Changgang) | 93276 | 3.1% | 3.4% (Δ +0.2 pp) | 车陂 (Chebei) | 77918 | 16.7% | 9.1% (Δ -7.6 pp) |
| 杨箕 (Yangji) | 71013 | 3.1% | 3.1% (Δ +0.0 pp) | 低涌 (Dichong) | 2752 | 15.9% | 5.6% (Δ -10.3 pp) |
| 三元里 (Sanyuanli) | 122190 | 2.9% | 2.9% (Δ +0.0 pp) | 黄村 (Huangcun) | 51488 | 15.9% | 9.8% (Δ -6.1 pp) |
| 萧岗 (Xiaogang) | 15319 | 2.8% | 2.8% (Δ +0.0 pp) | 蠕岗 (Leigang) | 16561 | 15.6% | 15.6% (Δ +0.0 pp) |
| 广州火车站 (Guangzhou Huochezhan) | 147541 | 2.7% | 2.7% (Δ +0.0 pp) | 江泰路 (Jiangtailu) | 61446 | 15.6% | 17.8% (Δ +2.2 pp) |
| 动物园 (Dongwuyuan) | 42666 | 2.6% | 2.6% (Δ +0.0 pp) | **东山口 (Dongshankou)** | **63102** | **15.5%** | **12.7% (Δ -2.8 pp)** |

Notes: Total OD column is the total number of OD flows that use a station as the origin or destination, and the parenthesis includes the percentile of this flow across all stations. Build Line 10 column represents the share of that station's origin-only trips for which shortest-path hops reduce > 1 when the network includes full Line 10 versus without Line 10. Line 10 w/o New Interchange (Δ) column represents the same metric with the counterfactual Line 10 that remove Dongshankou/Shuqianlu as a Line-10 stop. A bypass is created to connect neighbors of Dongshankou Station. Parenthesis indicates the difference with the full Line 10.

### 5.3 Reflections on Network Tradeoff Framework Friction in Existing Planning Models

Over the course of the Miaoqianzhijie–subway tradeoff debate, planning documents and local news reveal disparate planning logics that lean on place-based and spatial network-based frameworks, yet together they prove insufficient to address public sentiments and complex needs in the debate (see Table 2 for the project timeline). At the early stage of network planning, Line 10 was justified as taking on a system role to "improve subway network structure, facilitate regional transportation integration, and alleviate city center congestion" (NDRC, 2017). The subsequent back-and-forth over whether *Shuqianlu Station* should function as an interchange shows preliminary recognition of social complexity in nearby areas, yet improving overall network efficiency (GZDRC, 2018) was consistently prioritized over localized social and environmental concerns. Even when the demolition was cancelled eventually, Guangzhou Subway Company prioritize constructing Line 10 and skipping controversial areas, rather than taking time to develop a win-win plan for a specific station (Li, 2025). These considerations indicate that subway planning was primarily guided by a spatial network and systems-engineering framework.

Place-based considerations became most explicit when the specific demolition plan was announced (Yuexiu District Government, 2020), which was also the first time that many residents became aware of the concrete implications of the proposed changes. A similar place-based logic appeared in the environmental impact assessment for *Shuqianlu Station*, where impacts were estimated in terms of the number of households exposed to noise and air pollution at "sensitive points," defined mainly as residential buildings and schools within the demolition area. In both documents, the affected population is effectively defined as property owners within the demolition boundary, with no explicit mention of small businesses or members of the public who frequently visit the area or maintain social ties there.

When the demolition plan was confronted by public sentiments, Guangzhou Municipal Government and Guangzhou Subway Company again justified the station's location through two contrasting logics: on the one hand, elevating *Shuqianlu Station*'s role in the spatial network by emphasizing its contribution to relieving transfer burdens in the existing system; on the other, downplaying the place-based social value of *Miaoqianzhijie* by arguing that no officially designated historic buildings would be demolished. Local media, in turn, largely defended the street through broad place-based narratives of cultural preservation and collective memory, without explicitly tracing how these sentiments were anchored in specific social infrastructures and dense social connections, missing an opportunity to build a more nuanced, evidence-based foundation for mitigation or alternative designs. This gap also helps explain why the Guangzhou Subway Company underestimated community opposition: many non-local visitors had developed strong place attachments and relied on the street's social infrastructure to sustain their social relationships, extending the constituency of concern well beyond nearby residents.

From the perspective of the network tradeoff framework, this timeline illustrates not only a policy adjustment but also a misalignment between underlying logics: planning models that separately optimize the mobility network and regulate physical parcels, and public discourses that implicitly defend a street-level social network that these models rarely capture. The contribution of the network tradeoff framework is to make this misalignment explicit. It exemplifies how subway lines should be analyzed as spatial networks and historic streets as social networks within a single analytic frame, and that future planning practice should compare regional mobility gains and local social connectivity losses as explicit network tradeoffs, rather than treating them as technically quantified benefits on one side and "intangible" emotional claims on the other.

An explicit benefit of the network tradeoff framework is its ability to mobilize non-local activists who advocate for historic street preservation, bridge perspectives between planners and residents, and reframe preservation debates around social connectivity rather than place-based sentiment alone. For example, the social infrastructure map (Figure 3) helped initiate conversations with both local and non-local stakeholders by making visible the dense social ties supported by the street. Respondents appreciated seeing their everyday routines and local businesses represented, and the map guided oral history interviews with stakeholders, including non-local actors such as the retired director of the Yuexiu District Children's Palace, who recognized the street as a hub for family networks and community connections (Figure 5).

Additionally, both community members and planners found this framing of the issue as a network tradeoff between local social connectivity and regional mobility gains productive to move past polarized preservation debates. The traditional placed-based preservation narrative limits the solutions to preserving a few buildings, whereas the network perspective highlights key social infrastructure (a subset of small businesses) that sustains community relationships and shifts attention toward preserving infrastructures of social life.

More broadly, the process of constructing and communicating this narrative consolidated dispersed local knowledge into collective action. The findings were formalized in a report shared with the *Yuexiu District Planning and Natural Resource Bureau* (Figure 5). In 2022, Guangzhou Subway company paused the demolition plan for *Miaoqianzhijie* for further discussions and in 2025, *Shuqianlu Station* was permanently removed from Line 10. Although we cannot claim a causal link to the final decision, the *Miaoqianzhijie Research Group* helped align business owners, journalists, students, and residents, and helped "make legible" the street's social value through a combination of evidence and lived stories.

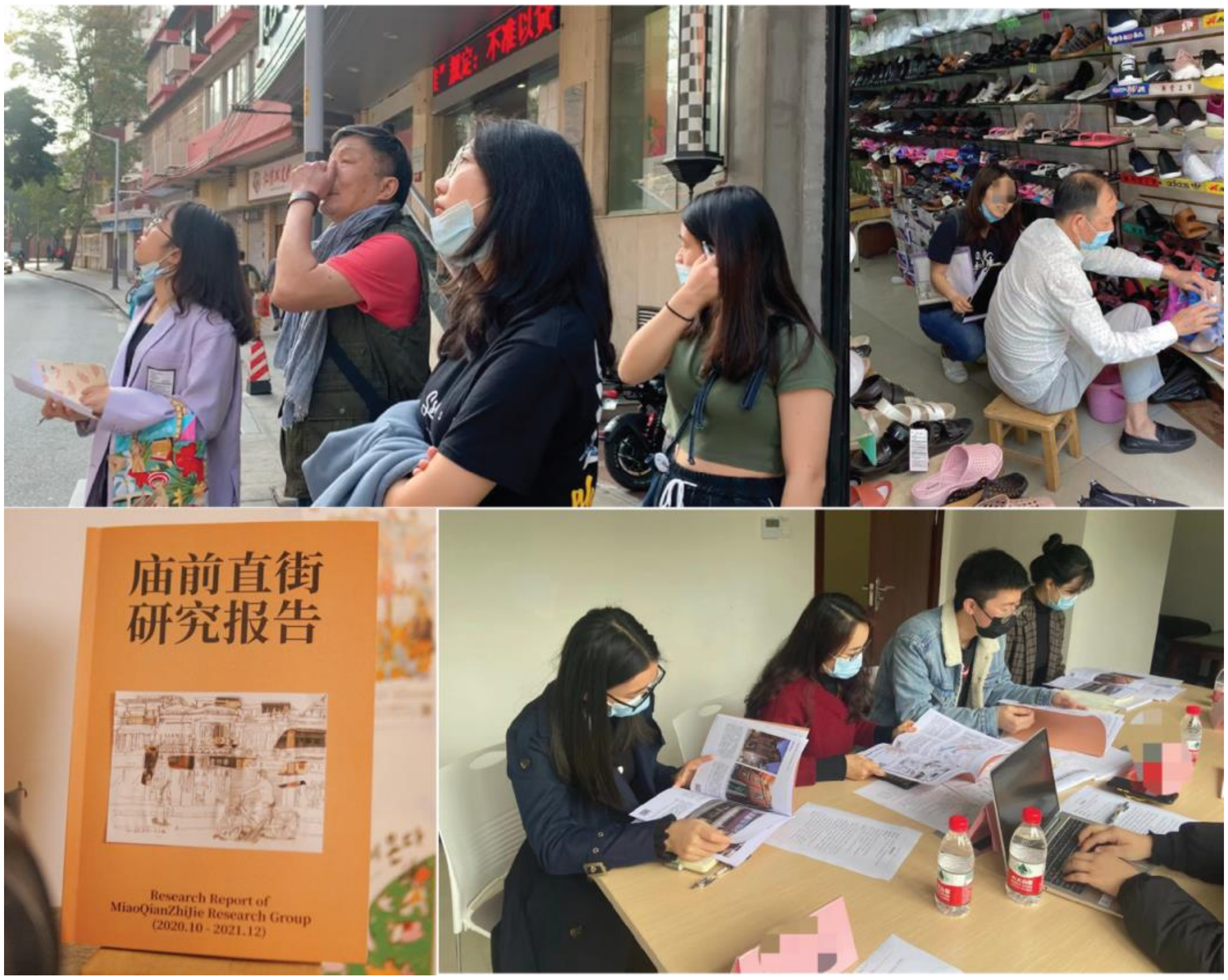

**Figure 5.** (top) Miaoqianzhijie Research Group volunteers conducting field work and oral history interviews; (bottom) Miaoqianzhijie Research Report submitted to Yuexiu District Planning and Natural Resource Bureau for review.

However, our conversations with municipal officers and a Guangzhou Subway Company informant highlight structural challenges in discussing network tradeoff, especially social connectivity, in infrastructure planning. Subway lines and stations are often designed or planned top-down and impacted residents' arguments focused on local experience are underrated in the cost and benefit analyses against the "public good" nature of the subway systems. Local governments also lack policy and legal frameworks to account for or remediate the loss of social ties and collective memories. These social factors are often lumped into a single, vague category of "social costs," which in our case was addressed primarily by compensating property owners and managing public sentiment in the media, rather than grappling with the underlying loss of social ties and collective memories. Traditional historical preservation schemes also fail to protect *Miaoqianzhijie* legally as it only covers built environments with architectural values and historical significance, but not those that support collective memories or social ties. After the controversy, a local government-led opinion draft to add collective memory as a new criterion for historical preservation in the legal framework was also pushed back due to the lack of clear standards to define and quantify collective memories (He & Fang, 2022).

The transient nature of social connectivity, which is constantly reshaped by evolving human dynamics and social infrastructure, also makes it hard to translate into a stable metric in planning models. For example, a year after the demolition controversy, our field volunteers found several local stores were already replaced by modern, fashionable, and higher-priced ones. Informants attributed these changes to rising rents and gradual gentrification, which they linked partly to the heightened visibility and foot traffic the controversy brought. At the same time, some long-term business owners reported considering relocation because the prolonged uncertainty around demolition made it difficult to plan or invest. Thus, the snapshot condition presented in this study could be outdated quickly as businesses, routines, and relationships reconfigure.

These challenges are not unique to Guangzhou. In the United States, research on NEPA and megaproject governance similarly finds that increasingly detailed environmental review has improved transparency and occasionally empowered marginalized groups, yet it has also lengthened project timelines, increased costs, and sometimes been strategically used by powerful stakeholders to shape outcomes (Flyvbjerg et al., 2003). Data and modeling constraints make it difficult to integrate detailed social network information into future-oriented accessibility or travel demand models, while social choice problems remain constant challenges on how to weigh dispersed regional mobility gains against concentrated local social losses (Pereira et al., 2017). While difficult, this case study shows that traditional planning models could run into unexpected roadblock if network tradeoffs are not deliberated. Thus, the solution is not only to recognize network tradeoffs conceptually, but also to develop practicable metrics, participatory procedures, and legal tools that can bring street-level social networks into the same decision frame as regional mobility networks, without paralyzing infrastructure delivery. This study is a preliminary attempt to visualize and deliberate these impacts side by side.

**Table 5.** Miaoqianzhijie-subway tradeoff debate timeline, anchored by signature planning documents and local news.

| **Timeline** | **Planning Documents / Local News** | **Implications on Shuqianlu Station / Miaoqianzhijie** |
|---|---|---|
| 2016-2017 Subway Network Planning | "Guangzhou Phase III Urban Rail Transit Construction Plan 2017-2023" was publicized and approved by Guangzhou and National | Line 10 was first mapped and *Shuqianlu Station* was first identified in the planned subway map as an interchange station. |
| | Development and Reform Commission (NDRC) in 2017 (NDRC, 2017). | |
| 2018-2019 Station Construction Planning | "Environmental Impact Assessment of Guangzhou Subway Line 10" was publicized (Guangzhou Subway, 2018). Guangzhou Development and Reform Commission (GZDRC) replied to Line 10 Feasibility Study (GZDRC, 2018) | *Shuqianlu Station* was discussed as an environmental sensitive location; GZDRC requested Guangzhou Subway Company to deliberate on the proposal to change *Shuqianlu Station* to a non-interchange station. |
| 2020-2021 Community Discontent | "Notice on Public Consultation for the Draft Compensation Scheme for Expropriation of Houses on State-Owned Land for the Guangzhou Rail Transit Line 10 Shuqianlu Station" (Yuexiu District Government, 2020) was publicized and raised media attention. | *Shuqianlu Station* was again described as a three line interchange station. The demolition area (including *Miaoqianzhijie*) for *Shuqianlu Station* was first mentioned and explicitly mapped. The demolition plan received intense public pushback. In a few weeks, Guangzhou Subway Company adjusted the demolition area to remove some institutional buildings but keep the north side of *Miaoqianzhijie* in the demolition plan. Demolition plan was paused for further evaluation. |
| 2022-2025 Quiet Phase | Local news reported in 2022 that the main construction vehicle has passed the *Shuqianlu Station* and in 2025, the new Line 10 route map did not include *Shuqianlu Station* (Li, 2025). | *Shuqianlu Station* was removed from Line 10 permanently. Demolition stopped. |

## 6 Conclusion

This study advances a network tradeoff framework for transportation planning by showing how a single project simultaneously reshapes connectivity across different networks. In terms of impact area, the demolition affects a highly localized social infrastructure system anchored in 89 small businesses (57 on the north side), where 278 recorded visits and 131 location-based memories reveal dense, place-based interactions, while the mobility benefits extend across the entire subway network through changes in OD flows. In terms of population, the social impacts are borne by those whose routines and relationships are tied to the street, particularly local residents and long-term users, among whom 62% report daily importance and 55.7% report support for social ties, whereas mobility benefits accrue more broadly to metro users across the system, with uneven gains across stations. In terms of empirical measurement, the social network is characterized by concentrated, high-attachment ties, with 71% of visits involving companions and 64% of local stakeholders reporting mutual help, and limited substitutability, as 39% of activities would be dropped and only 26% could be relocated nearby if the street is removed. In contrast, the mobility network is evaluated through changes in travel efficiency and routing, with relatively modest benefits for the closest station, *Dongshankou* (5.3% of flows benefiting from distance reduction and 15.5% gaining additional routes), and broader but uneven improvements across the network. In terms of equity considerations, these results reveal a fundamental asymmetry: social costs are concentrated among a smaller, highly dependent population with long-term, place-based attachments, while mobility gains are diffuse, incremental, and distributed across a wider set of users. By aligning these impacts within a single framework, the study highlights how infrastructure projects can prioritize certain forms of connectivity over others, making visible the uneven distribution of benefits and costs that is often obscured in "public good" narratives of subway expansion.

We recognize several limitations in our approach. First, our measurement of social impact could be more objective and in-depth. For instance, we did not quantify the availability of equivalent amenities nearby or measure "replaceability" using comparative POI access distances. Our perception-based survey methods may overstate the uniqueness of *Miaoqianzhijie* due to anchoring bias. Respondents may not have fully considered functional substitutes. Additionally, we made strong assumptions about the disruption of social ties if small businesses were displaced, without tracking how those networks evolve or persist beyond the immediate location. A more robust analysis would include social network mapping or longitudinal observation of relocation and adaptation, which were not feasible within our time constraints or institutional limitations.

More fundamentally, our social network analysis lacks a baseline for evaluating the relative size, strength, and substitutability of place-based social infrastructure. Without such a benchmark, it remains difficult to assess how exceptional *Miaoqianzhijie* is or how easily its social functions can be reproduced elsewhere. Future research could address this gap by developing baseline models of "expected replaceability," for example by estimating access to functionally similar POIs within spatial or network proximity, or by examining how similar populations re-anchor their routines across alternative locations. Extending this approach to the street level could also help evaluate how replicable a given mix of social infrastructure is, providing a more systematic basis for comparing social impacts across cases. Such baseline models could also inform practical strategies for mitigating social disruption and rebuilding social ties after construction. For instance, if certain businesses are identified as low-substitutability, high-attachment nodes, planners could prioritize their preservation in situ, phased relocation nearby, or coordinated clustering in designated receiving areas to maintain existing interaction patterns.

Second, our survey approach likely overrepresents residents who feel strongly about the site. While this is typical for participatory research, it introduces selection bias. Not all locals or frequent visitors have strong attachments to the street. Consequently, our results should be interpreted as a detailed portrait of the networks and attachments of engaged users, rather than a statistically representative characterization of all residents in the impact area. Our aim is to demonstrate a network-based method for foregrounding the social connectivity of a historic street, not to estimate citywide prevalence of specific attitudes.

Similarly, our OD hop-based analysis is a stylized representation of spatial benefits. Our OD analysis assumes that travel demands to the existing subway destinations would remain unchanged with the introduction of Line 10. However, new stations can stimulate previously impossible trips or offer more transit combinations, potentially benefiting locals more than our analysis suggests. Conversely, we did not account for the time and cost of transferring between subway lines when calculating hops or possible routes, which may have led to an overestimation of the station's spatial benefits. Due to limited data and information about the proposed changes, we cannot model future demand or induced trips under alternative station designs; instead, we use existing flows to highlight the uneven distribution of potential gains across the network and to situate local social costs within a broader but partial picture of regional accessibility.

Despite these limitations, this study contributes to a growing body of research that seeks to reframe infrastructure conflicts through the network tradeoff framework. Balancing these opposing outcomes requires new tools and frameworks; our study is just the first step to recognize multiple types of connectivity and the differentiated impacts of infrastructure on diverse populations.

## Acknowledgments

We would like to acknowledge Clio Andris for supporting the first author’s research work, Yuxiang Zhao for supporting graphic visualization, and Miaoqianzhijie Research Group volunteers Qian Cai, Ninghui Wang, Xiayi He, Shihui Liu, Chunchun Cai, Sanshi Yang for collecting data, creating promotional materials, writing social media stories, and supporting background research. We would also like to thank Canton Gaai Fong Cing & Book Docent (广州街坊情/BOOK 導賞) for supporting our fieldwork.

Changdong Ye would like to acknowledge funding support from 2026 General Project of the Guangdong Provincial Philosophy and Social Sciences Planning Project: “Age-Friendliness Assessment and Response Strategies for Public Spaces in Older Residential Communities under the Context of Delayed Retirement” (Grant No. GD26CSH14).

## Author contribution

The authors confirm their contribution to the paper as follows: Writing and Conception: Xiaofan Liang; Visualization and Analysis: Xiaofan Liang; Data Collection, Interview, and Community Engagement: Yun Tian, Manying Lyu, Lu Chen, Xiaofan Liang, Changdong Ye.